\documentclass[english,letterpaper,twocolumn,showpacs,prl,aps]{revtex4}
\usepackage{graphicx}
\usepackage{amssymb}

\makeatletter

\large  

\input epsf

\makeatother

\usepackage{babel}
\makeatother
\begin{document}

\title{Brownian motion of the electron and the Lamb shift at finite temperature}

\author{Eugene B. Kolomeisky}

\affiliation
{Department of Physics, University of Virginia, P. O. Box 400714,
Charlottesville, Virginia 22904-4714, USA}

\begin{abstract}
By enhancing electron position fluctuations, equilibrium electromagnetic radiation modifies the potential for an electron in a Hydrogen atom.  This can have significant effects for weakly bound states and especially at finite temperature.  This implies a $2\%$ correction to Bethe's value for $2S_{1/2}-2P_{1/2}$ Lamb shift for weak fluctuations, but the effect is an order of magnitude larger for strong fluctuations where it provides direct measure of the proton diameter.    
\end{abstract}

\pacs{31.30.jf, 05.40.-a, 05.40.Jc, 03.65.Sq}

\maketitle

In 1948, following Bethe's classic calculation \cite{Bethe} of the experimentally observed energy splitting of the $2S_{1/2}$ and $2P_{1/2}$ levels of Hydrogen, the celebrated Lamb shift \cite{Lamb}, Welton \cite{Welton} presented an elegant semiclassical explanation of the effect in terms of fluctuations of the electron position superimposed on its orbital motion.  The Brownian motion of the electron is due to the coupling to zero-point fluctuations of the electromagnetic field.   Welton observed that this generates an effective repulsive barrier at the origin which he treated via first-order perturbation theory.  The goal of this paper is to point out that the assumption that this induced potential is weak is not satisfied as well as one might expect.  This leads to a small correction to the Bethe value of the  $2S_{1/2}-2P_{1/2}$ level splitting.  The effect is an order of magnitude larger for strong electron fluctuations  which will be the case for a sufficiently weakly bound electron, and especially at finite temperature.  Since these findings have non-relativistic origin and Welton's analysis, while reproducing the bulk of the Lamb shift, is unaffected by  the many-particle and relativistic effects \cite{CT},  the full quantum electrodynamics treatment \cite{LL4} is unnecessary. 

Fluctuations of the electron position induce changes both in the kinetic and potential energy.  The former can be ignored because they are the same for both the $S$ and $P$ states; only the change in the potential energy $U(\textbf{r})$ needs to be considered.  When the electron trajectory is perturbed so that $\textbf{r}\rightarrow \textbf{r}+ \delta\textbf{r}$ where $\delta\textbf{r}$ is an isotropic fluctuation with zero average $<\delta\textbf{r}>=0$ and finite mean-square displacement $<\delta\textbf{r}^{2}>$, the potential energy felt by the electron will be given by \cite{Welton}
\begin{equation}
\label{perturbation}
<U(\textbf{r}+\delta\textbf{r})>=\exp \left (\frac{1}{6}<\delta\textbf{r}^{2}>\nabla^{2}\right )U(\textbf{r})
\end{equation}
If the proton would be a point charge $e$, then $U(\textbf{r})=-e^{2}/r$ and the extra induced potential felt by the electron would be \cite{Welton}   
\begin{equation}
\label{delta-function_potential}
\delta U(\textbf{r})\simeq \frac{2\pi}{3}e^{2}<\delta \textbf{r}^{2}> \delta(\textbf{r})
\end{equation}
Such three-dimensional delta-function potential, in contrast to its one-dimensional counterpart, has a formal meaning and can only be viewed as a perturbation and only to first order.  The true potential at small distances is a barrier of width $r_{0}$, the proton radius, and height $\delta U \simeq e^{2}<\delta \textbf{r}^{2}>/r_{0}^{3}$;  whether this barrier can be regarded as weak or strong requires separate investigation.  In order to do so properly, we will model the proton by a uniformly-charged ball of radius $r_{0}$.  Then for $r>r_{0}$ the potential energy of the electron is $U(\textbf{r})= -e^{2}/r$ while for $r\leqslant r_{0}$ it is given by
\begin{equation}
\label{inner_well}
U(\textbf{r})=-\frac{e^{2}}{2r_{0}}\left (3-\frac{r^{2}}{r_{0}^{2}}\right )
\end{equation} 
As a result the electron experiences an attraction that is weaker than $-e^{2}/r$ at small distances, which \textit{increases} its energy.  The corresponding energy shift is quadratic in the proton radius $r_{0}$ \cite{LL3} regardless of the model of the charge distribution inside the proton.    Since $r_{0}/a_{B}=1.66 \times 10^{-5}$ \cite{Codata}, where $a_{B}=\hbar^{2}/me^{2}$ is the Bohr radius, the finite size correction to the energy due to (\ref{inner_well}) will be neglected (see below) compared to the fluctuation effect.   

Substituting (\ref{inner_well}) into (\ref{perturbation}) we observe that due to the position fluctuations of the electron the potential energy acquires an extra contribution given exactly by 
\begin{equation}
\label{extra}
\delta U(\textbf{r})=\frac{e^{2}<\delta \textbf{r}^{2}>}{2r_{0}^{3}}
\end{equation}              
which is a constant height barrier operating at $r\leqslant r_{0}$.  The strength of such a barrier can be judged by forming the dimensionless combination 
\begin{equation}
\label{barrier_strength}
\lambda=\sqrt{\frac{2m \delta Ur_{0}^{2}}{\hbar^{2}}}=\left (\frac{<\delta \textbf{r}^{2}>}{a_{B}^{2}}\frac{a_{B}}{r_{0}}\right )^{1/2}
\end{equation}
The consequence is that because $a_{B}/r_{0}\gg 1$, the condition of weak electron fluctuations $<\delta\textbf{r}^{2}> /a_{B}^{2}\ll 1$ alone does not guarantee the weakness of the induced barrier (\ref{extra}).   The crossover between the regimes of weak $\lambda\ll1$ and strong $\lambda\gg1$ barriers occurs at $<\delta \textbf{r}>^{2}/a_{B}^{2}\simeq r_{0}/a_{B}\approx10^{-5}$, i.e. when position fluctuations are small. 

We thus face a Coulomb problem modified at small distances by the presence of the barrier (\ref{extra});  principally the $S$ states are affected by the barrier.  The upper part of the $S$-state energy spectrum of a generic modified Coulomb problem is described by the Rydberg formula
\begin{equation}
\label{Rydberg}
E_{n}(\mu)=const-\frac{1}{2(n-\mu)^{2}}~~~(a.u.)
\end{equation} 
where we employed atomic units (\textit{a.u.}), the $const$ is common to the $S$ and $P$ states and $\mu$ is the quantum defect \cite{KT}.   The latter represents convenient dimensionless measure of experimental splitting between the $S$ and $P$ levels determined by the difference of (\ref{Rydberg}) evaluated at finite and zero $\mu$.   The quantum defect can be expressed in terms of the $S$-wave scattering length $a_{s}$ of the $r\leqslant r_{0}$ potential \textit{only} \cite{KT}.  In the present case of the constant height potential barrier (\ref{extra}) the scattering length is given by 
\begin{equation}
\label{scattering_length}
a_{s}= r_{0}\left (1-\frac{\tanh\lambda}{\lambda}\right )
\end{equation}
Since $a_{s}\leqslant r_{0}$ for all $\lambda$ and $r_{0}\ll a_{B}$, the scattering length also satisfies the condition $a_{s}\ll a_{B}$ (which is also true for any reasonable model of charge distribution inside the proton).   This is the range of applicability of the perturbation theory in $a_{s}/a_{B}$, where the quantum defect is given by \cite{Deser,KT}
\begin{equation}
\label{qdefect_general}
\mu= - \frac{2a_{s}}{a_{B}}
\end{equation}

In the limit of a weak barrier $\lambda\ll1$ (with $\lambda$ given by Eq.(\ref{barrier_strength})) the scattering length (\ref{scattering_length}) approaches a limit that is independent of the proton radius $r_{0}$: $a_{s}=r_{0}\lambda^{2}/3=<\delta \textbf{r}^{2}>/3a_{B}\ll r_{0}$.  The corresponding quantum defect  
\begin{equation}
\label{qdefect_Welton}
\mu_{0}=-\frac{2}{3}\frac{<\delta \textbf{r}^{2}>}{a_{B}^{2}}
\end{equation}
directly measures electron's mean-square fluctuation.  This conclusion is equivalent to the result of Welton \cite{Welton} that follows from the $|\mu|/n\ll 1$ expansion of the Rydberg formula (\ref{Rydberg});  this is the regime of applicability of the perturbation theory in the induced potential (\ref{delta-function_potential}). 

In the large-barrier limit $\lambda\gg1$ the scattering length $a_{s}$ (\ref{scattering_length}) approaches the size of the proton $r_{0}$ with the quantum defect
\begin{equation}
\label{qdefect_strong}
\mu_{\infty}=-\frac{2r_{0}}{a_{B}}=-3.32\times10^{-5}
\end{equation}  
directly measuring the negative of the proton diameter.  We note that the weak (\ref{qdefect_Welton}) and strong (\ref{qdefect_strong}) barrier results are insensitive to the model of the charge distribution inside the proton. 

We now proceed to a calculation of the mean-square fluctuation $<\delta\textbf{r}^{2}>$ which will tell us what regime the real system may belong to.  Consider a non-relativistic ($|\dot{\textbf{r}}|\ll c$) harmonically bound electron of mass $m$ driven by a fluctuating time-dependent electric field $\textbf{E}(t)$.  The force on the electron corresponding to this field is  $\textbf{f}=-e\textbf{E}$ and the equation of motion for the electron -- including radiation damping -- is \cite{LL2}     
\begin{equation}
\label{eqmotion}
m\ddot {\textbf{r}}-\frac{2e^{2}}{3c^{3}}\dddot{\textbf{r}}+m\omega_{0}^{2}\textbf{r}=\textbf{f}(t)=-e\textbf{E}(t) 
\end{equation}     
where $\omega_{0}$ is an oscillation frequency representing the frequency of motion in the Bohr orbit.  This equation differs from that employed by Welton in one crucial respect:

As written, Welton's equation of motion of a free ($\omega_{0}=0$) electron $m\ddot \textbf{r}= -e\textbf{E}(t)$ with $\textbf{E}(t)$ representing the fluctuating equilibrium electric field contradicts the fluctuation-dissipation theorem (FDT) \cite{FDT,LL5} since it contains fluctuations but no dissipation.  It was indeed later recognized by Callen and Welton \cite{FDT} that application of the FDT to Eq.(\ref{eqmotion}) leads to the Planck formula for the spectral energy density of the radiation.  Thus the radiation damping term $-(2e^{2}/3c^{3})\dddot{\textbf{r}}$ in the equation of motion (\ref{eqmotion}) is both appropriate and necessary.  

Introducing a spectral decomposition of the fluctuating quantities according to the conventions of Ref. \cite{LL5}, $\textbf{r}(t)=\int_{-\infty}^{\infty}\textbf{r}_{\omega}e^{-i\omega t}\frac{d\omega}{2\pi}$, the equation of motion (\ref{eqmotion}) can be written in the spectral form as $\textbf{r}_{\omega}=\mathcal{A}(\omega)\textbf{f}_{\omega}$ where $\mathcal{A}(\omega)=\mathcal{A'(\omega)}+i \mathcal{A''(\omega)}$ is the generalized susceptibility:
\begin{equation}
\label{susceptibility}
\frac{1}{\mathcal{A}(\omega)}= m(\omega_{0}^{2}- \omega^{2})-i\frac{2e^{2}\omega^{3}}{3c^{3}}
\end{equation}
This equation is applicable for frequencies that are significantly smaller than the inverse time of travel of light through a distance of the order of the classical radius of the electron, $\omega \ll mc^{3}/e^{2}$;  ignoring this constraint of internal consistency of classical electrodynamics is known to lead to physically absurd results \cite{LL2}.  For our problem this condition is practically irrelevant because quantum effects come into play at frequencies of the order $mc^{2}/\hbar$ which is $\alpha^{-1}=\hbar c/e^{2}=137$ times smaller than the classical limit \cite{LL2}.     

The generalized susceptibility (\ref{susceptibility}) determines, via the FDT, the mean-square displacement of the electron \cite{LL5}: 
\begin{equation}
\label{msdisplacement_general}
<\delta \textbf{r}^{2}>=\frac{3\hbar}{\pi}\int_{0}^{\infty}\mathcal{A}''(\omega)\coth\frac{\hbar\omega}{2T}d\omega
\end{equation}       
where the factor of $3$ accounts for the three components of the displacement vector $\textbf{r}$ and $T$ is a temperature.  In the neutral limit $e\rightarrow 0$ the integral (\ref{msdisplacement_general}) can be evaluated with the result 
\begin{equation}
\label{Bloch}
<\delta \textbf{r}^{2}>=\frac{3\hbar}{2m\omega_{0}}\coth\frac{\hbar\omega_{0}}{2T}
\end{equation}
that can be recognized as Bloch's formula for the mean-square displacement of an undamped oscillator \cite {LL5}.  

For a free electron $\omega_{0}=0$, and at $T=0$ Eqs.(\ref{susceptibility}) and (\ref{msdisplacement_general}) lead to the result
\begin{equation}
\label{msdisplacement_zeroT_free}
<\delta \textbf{r}^{2}>=\frac{2\alpha^{3}}{\pi}a_{B}^{2}\int_{0}^{\infty}\frac{d\omega}{\omega+(2e^{2}/3mc^{3})^{2}\omega^{3}}
\end{equation}
This expression differs from Welton's only in the presence of the $(2e^{2}/3mc^{3})^{2}\omega^{3}$ term in the denominator.  The frequency integral is logarithmically divergent at the lower limit;  for a bound electron the effective infrared cutoff is $\omega_{0}$ \cite{Welton}.  In contrast to Welton's result, the integral in (\ref{msdisplacement_zeroT_free}) is ultraviolet convergent;  the effective frequency cutoff is set by $mc^{3}/e^{2}$, the range of applicability of classical electrodynamics.  Welton recognized the necessity for a physical ultraviolet cutoff which he set at a frequency of the order $mc^{2}/\hbar$ where quantum-mechanical effects come into play.  Therefore it is appropriate to ignore the $(2e^{2}/3mc^{3})^{2}\omega^{3}$ term in the denominator in (\ref{msdisplacement_zeroT_free}) and instead set the upper integration cutoff at $mc^{2}/\hbar$.  This leads to Welton's result for the mean-square displacement of a bound electron
\begin{equation}
\label{msdisplacement_zeroT_bound }
\frac{<\delta \textbf{r}^{2}>}{a_{B}^{2}}\approx\frac{2\alpha^{3}}{\pi}\int_{\omega_{0}}^{\frac{mc^{2}}{\hbar}}\frac{d\omega}{\omega}=\frac{2\alpha^{3}}{\pi}\ln\frac{mc^{2}}{\hbar\omega_{0}}=\frac{2\alpha^{3}}{\pi}\ln\frac{n^{3}}{\alpha^{2}}
\end{equation}
which has logarithmic accuracy.  Comparing the $\omega_{0}$-dependence here and in Eq.(\ref{Bloch}) at $T=0$ we conclude that the fluctuations of charged oscillator are dramatically suppressed.  In the last step in (\ref{msdisplacement_zeroT_bound }) we employed 
\begin{equation}
\label{Bohr_frequency}
\omega_{0}=\frac{e^{2}}{\hbar a_{B}n^{3}}
\end{equation}
(Bohr's correspondence principle) for the frequency of motion in Bohr's $n$-th orbit applicable for $n\gg1$.  For the $2S$ state and reasonable cutoffs the logarithm in (\ref{msdisplacement_zeroT_bound }) is approximately $8$ \cite{Bethe} and then Eq.(\ref{msdisplacement_zeroT_bound })  tells us that at zero temperature the root-mean-square fluctuation of the electron in the Bohr orbit is about $700$ times smaller than the Bohr radius.  Even though the mean-square displacement grows with the principal quantum number, the $n$-dependence is weak:  for $n=50$ the root-mean-square fluctuation is only about $1.5$ times larger than for $n=2$.   For the same $2S$ state substitution of (\ref{msdisplacement_zeroT_bound }) into (\ref{qdefect_Welton}) leads to the quantum defect $\mu_{0}=-1.32 \times 10^{-6}$ whose magnitude is $25$ times smaller than the strong barrier result (\ref{qdefect_strong}).   This however assumed that the induced barrier can be regarded as weak.  In order to verify whether this is the case we compute the effective strength parameter (\ref{barrier_strength}) with the result $\lambda\approx0.35$ which is smaller than unity (but not by a lot).  Then a recalculation of the quantum defect (\ref{qdefect_general}) with the scattering length evaluated according to Eq.(\ref{scattering_length}) with $\lambda=0.35$ gives $\mu=-1.29 \times 10^{-6}$ which is $2.27 \%$ different from Bethe's value.  This $2.27\%$ difference has the same order of magnitude as the contribution into the Lamb shift due to vacuum polarization \cite{LL4} and is two orders of magnitude larger than the leading finite-size correction contributing  an amount of the order $-(r_{0}/a_{B})^{2}$ \cite{LL3} into the quantum defect.  We also note that what we believe is the correct \textit{non-relativistic} $2S$ quantum defect ($\mu=-1.29 \times 10^{-6}$) depends only weakly on the assumption of a uniform charge distribution inside the proton.

It would be important to verify whether this effect could resolve the ongoing controversy regarding the size of the proton because presently accepted value of the proton radius \cite{Codata} is extracted from precision spectroscopy of atomic Hydrogen.  However supposedly more accurate measurement involving muonic Hydrogen gives approximately a $4\%$ smaller value of the radius \cite{size}.  Our analysis is not applicable to the case of muonic Hydrogen where the bulk of the effect is due to polarization of electron vacuum \cite{LL4}:  the muon Bohr radius $a_{B}/207$ is within the $a_{B}/137$ range where polarization of the electron vacuum substantially modifies the Coulomb law.  Thus the muon experiences an attraction that is stronger than $-e^{2}/r$ at small distances which \textit{decreases} its energy.

It is also important to ask if it is possible to observe much larger deviations from the Bethe-Welton value (\ref{qdefect_Welton}), specifically the impenetrable barrier limit (\ref{qdefect_strong}).  A generalization of our analysis to the case of a Hydrogenic ion of atomic number $Z\ll137$ and atomic mass $A$ predicts that  with logarithmic accuracy this amounts to multiplication of Eq.(\ref{barrier_strength}) by $Z^{1/2}A^{-1/6}$.  For sodium ion ($Z=11$, $A=23$) this increases the effective strength parameter (\ref{barrier_strength}) by a factor of $2$.   However, significantly larger values of $\lambda$ can be realized at finite temperature and $n\gg1$ due to enhancement of the electron fluctuations.

For a free electron ($\omega_{0}=0$), a counterpart to Eq.(\ref{msdisplacement_general}) was given by Moore \cite{Moore} who concluded that thermal fluctuations have negligible effect on the $2S_{1/2}-2P_{1/2}$ Lamb shift.  Although we agree with such an assessment, Moore's account of the electron binding was incorrect;  we also disagree with his choice for the ultraviolet cutoff. 

At finite temperature the mean-square displacement (\ref{msdisplacement_general}) can be approximately computed by setting $\coth x\approx1/x$ for $x<1$ and $\coth x\approx 1$ for $x\geqslant 1$, which separates the fluctuations into the classical ($\omega \ll T/\hbar$) and quantum ($\omega \gg T/\hbar$) ranges \cite{LL5} :
\begin{equation}
\label{msdisplacement_finiteT}
<\delta\textbf{r}^{2}>\approx\frac{6T}{\pi}\int_{0}^{\frac{2T}{\hbar}}\frac{\mathcal{A}''(\omega)}{\omega}d\omega+ \frac{3\hbar}{\pi}\int_{\frac{2T}{\hbar}}^{\frac{mc^{2}}{\hbar}}\mathcal{A}''(\omega)d\omega
\end{equation}
Assuming that the temperature is high enough  ($\hbar\omega_{0}\ll T$) allows us to calculate both integrals:

In the first integral the upper limit can be replaced with infinity and the integral, with the help of the Kramers-Kronig relationship \cite{LL5}, is evaluated to $(\pi/2)\mathcal{A}'(0)=(\pi/2)\mathcal{A}(0)=\pi/(2 m\omega_{0}^{2})$.  For the second integral we can neglect $\omega_{0}$ relative to $\omega$ in (\ref{susceptibility}), so that $\mathcal{A}''(\omega)= 2e^{2}/3m^{2}c^{3}\omega$.  Then the mean-square displacement of the electron will be given by
\begin{eqnarray}
\label{msdisplacement_highT}
<\delta\textbf{r}^{2}>&\approx&\frac{3T}{m\omega_{0}^{2}}+ \frac{2\alpha^{3}}{\pi}a_{B}^{2}\ln\frac{mc^{2}}{T}\nonumber\\
&\approx&3Tn^{6}+ \frac{2\alpha^{3}}{\pi}\ln\frac{1}{T\alpha^{2}}~~~(a.u.)
\end{eqnarray}   
where in the second representation we employed Eq.(\ref{Bohr_frequency}).  The second terms in (\ref{msdisplacement_highT}) give the contribution of the quantum fluctuations while the first terms can be recognized as the classical mean-square displacement of the position of a three-dimensional harmonic oscillator (same as the $T\gg\hbar\omega_{0}$ limit of the Bloch formula (\ref{Bloch})).  

Comparing Eqs.(\ref{msdisplacement_zeroT_bound }) and (\ref{msdisplacement_highT}) we note a stronger divergence of the mean-square displacement in the latter case as the free-electron limit $\omega_{0}\rightarrow 0$ (or $n\rightarrow \infty$) is approached.  It is clear that the mean-square displacement cannot grow with $n$ without bound as predicted by Eq.(\ref{msdisplacement_highT}).  Indeed, the concept of the orbit retains its meaning only as long as the root-mean-square displacement is much smaller than the orbital size $n^{2}(a.u.)$.  This leads to the constraint $Tn^{2}\ll1(a.u)$.  It seems plausible that the states satisfying $Tn^{2}\gtrsim 1(a.u.)$ will be destroyed by thermal fluctuations;  this means that at room temperature $T\approx 10^{-3}a.u.$ the $n\gtrsim30$ states are unobservable. 

Moreover, one has to make sure that during an experiment the state in question does not decay, which requires that its lifetime \cite{Gallagher}   $n^{2}/(\alpha^{3}T)(a.u.)$ is significantly longer than the Kepler period $n^{3}(a.u)$, thus implying $n\ll\alpha^{-3}T^{-1}(a.u.)$.  However this condition is in practice irrelevant since it cannot compete for realistic temperatures with the requirement that the orbit is well-defined ($Tn^{2}\ll1(a.u.)$). Combining the $Tn^{2}\ll 1(a.u.)$ constraint with that of high-temperature limit $T\gg\hbar\omega_{0}$ with $\omega_{0}$ given by (\ref{Bohr_frequency}) provides us with a range of principal quantum numbers to which Eq.(\ref{msdisplacement_highT}) is applicable
\begin{equation}
\label{range}
T^{-1/3}\ll n \ll T^{-1/2}~~~(a.u.)
\end{equation}   
At room temperature this gives a range of $n$ between $10$ and $30$;  at 40 K the range is between $30$ and $100$.  With these values the mean-square displacement of the electron is dominated by the $3Tn^{6}$ term of (\ref{msdisplacement_highT}).        

The strong-barrier result (\ref{qdefect_strong}) should be observable in the regime of large electron fluctuations as described by Eqs.(\ref{msdisplacement_highT}) and (\ref{range}).  In practice we expect it to be valid in a much wider range of parameters because, as was already mentioned, the crossover between the regimes of weak and strong barriers occurs when the mean-square position fluctuation $<\delta r>^{2}/a_{B}^{2}\approx 10^{-5}$ is rather small.

The results described above have their origin in the divergent behavior of the mean-square fluctuation of the free electron.  Deeper understanding of what that means can be achieved by consideration of the correlation between displacements of a free ($\omega_{0}=0$) electron at different moments of time, which are measured by the correlation function $\mathcal{C}(t)=<\left (\textbf{r}(t)-\textbf{r}(0)\right )^{2}>$, again determined by the FDT \cite{FDT,LL5}:
\begin{equation}
\label{cfunction_general}
\mathcal{C}(t)=\frac{6\hbar}{\pi}\int_{0}^{\infty} \mathcal{A}''(\omega)\ (1-\cos\omega t)\coth\frac{\hbar \omega}{2T}d\omega
\end{equation} 
which at zero temperature and in the long time limit $\ln(mc^{2}t/\hbar)\gg1$ can be evaluated as 
\begin{equation}
\label{cfunction_zeroT_final}
\mathcal{C}(t)\approx \frac{4\alpha^{3}}{\pi}a_{B}^{2}\ln\frac{mc^{2}t}{\hbar}
\end{equation}
This implies that the electron executes a very slow sub-diffusive Brownian motion.

At finite temperature the correlation function $\mathcal{C}(t)$ can be computed in the $Tt/\hbar \gg1$ limit with the result 
\begin{equation}
\label{cfunction_finiteT}
\mathcal{C}(t)\approx \frac{4Te^{2}}{m^{2}c^{3}}t\equiv 6Dt,~~~~D=\frac{2Te^{2}}{3m^{2}c^{3}}\end{equation}
This can be recognized as the standard diffusion motion with diffusion constant $D$ whose physical origin can be understood by observing that the result (\ref{cfunction_finiteT}) is purely classical.  Therefore the Einstein relation between the diffusion constant $D$ and static mobility $\mathcal{B}'(0)$ must hold thus implying a very small mobility $\mathcal{B}'(0)=D/T=2e^{2}/(3m^{2}c^{3})$.  Since the frequency-dependent mobility $\mathcal{B}(\omega)=\mathcal{B}'(\omega)+i\mathcal{B}''(\omega)$ is the coefficient of proportionality between the spectral components of the electron's terminal velocity and the external force, we have $\mathcal{B}'(\omega)=\omega\mathcal{A}''(\omega)$.  Employing Eq.(\ref{susceptibility}) with $\omega_{0}=0$, and taking the static $\omega\rightarrow 0$ limit we again find $\mathcal{B}'(0)=D/T=2e^{2}/(3m^{2}c^{3})$.  Therefore we conclude that the free electron mobility and associated diffusion constant have their origin in the radiation damping.

We thank T. F. Gallagher, I. Shlosman and J. P. Straley for valuable comments.  This work was supported by US AFOSR Grant No. FA9550-11-1-0297.

\end{document}